\documentclass[preprint,journal]{vgtc}            

\onlineid{0}

\preprinttext{}

\vgtccategory{Research}

\vgtcpapertype{evaluation}

\title{Color, Gender, and Bias: Examining the Role of Stereotyped Colors in Visualization-Driven Pay Decisions}

\author{%
  \authororcid{Florent Cabric}{0000-0002-9326-9441},
  \authororcid{Margret Vilborg Bjarnadottir}{0000-0003-2955-1992}, and 
  \authororcid{Petra Isenberg}{0000-0002-2948-6417},
}

\authorfooter{
  \item
  	Josiah Carberry is with Brown University.
  	E-mail: jcarberry@example.com
  \item
  	Ed Grimley is with Grimley Widgets, Inc.
  	E-mail: ed.grimley@example.com.

  \item Martha Stewart is with Martha Stewart Enterprises at Microsoft
  Research.
  	E-mail: martha.stewart@example.com.
}

\abstract{%
  We investigate the impact of stereotyped gender-color associations in a visualization-driven decision-making task. 
  In the context of gender data visualization, the well-known ``\textcolor{W-Stereo}{pink} for girls and \textcolor{M-Stereo}{blue} for boys'' color assignment is associated with stereotypes that could bias readers and decision-makers. 
  Understanding the effects of using stereotyped colors in visualizations for decision-making can help designers better choose colors in stereotype-prone contexts. 
  We therefore explore the potential impact of stereotyped colors on compensation decision-making 
  through two crowdsourced experiments. In these experiments, we evaluate how the association of color with gender (stereotyped vs non-stereotyped) affects 
  the user's allocation decisions 
  in the context of salary adjustments. Our results indicate that explicit expression of the color-gender associations, in the form of a legend on the data visualization, leads to in-group favoritism. However, in the absence of a legend, this in-group favoritism disappears, and a small effect of non-stereotyped colors is observed.   
  A free copy of this paper with all supplemental materials is available at \url{https://osf.io/d4q3v/?view_only=22b636d6f7bb4a7991d9576933b3aaad}.
}


\graphicspath{{figs/}{figures/}{pictures/}{images/}{./}} 

\usepackage{lipsum}                    
\usepackage{mwe}                       
\usepackage{comment}
\usepackage{microtype}                 
\PassOptionsToPackage{warn}{textcomp}  
\usepackage{textcomp}                  
\usepackage{mathptmx}                  
\usepackage{times}                     
\usepackage{cite}                      
\usepackage{tabu}                      
\usepackage{booktabs}                  
\usepackage{soul}
\usepackage[dvipsnames,svgnames]{xcolor}
\usepackage{academicons}
\usepackage{tikz} 
\definecolor{lime}{HTML}{A6CE39}
\usepackage[nointegrals]{wasysym} 
\usepackage{multirow}
\usepackage{svg}
\usepackage{expdlist}
\usepackage{marvosym}
\usepackage{pdfpages}
\usepackage{amssymb}
\usepackage{flushend}
\usepackage{quoting}
\usepackage{framed}
\usepackage{xspace}
\usepackage{multirow}
\usepackage{graphicx}
\usepackage{subcaption}
\usepackage{enumitem} 
\setlist{noitemsep}

\newcommand{\ie}{i.e.,\@\xspace}
\newcommand{\eg}{e.g.,\@\xspace}
\newcommand{\etal}{et al.\@\xspace}

\newcommand{\todo}[1]{{\color{red}\bf{TODO: #1}\normalfont}}

\definecolor{W-Stereo}{HTML}{ff339c}
\definecolor{M-Stereo}{HTML}{3a33ff}
\definecolor{M-NonStereo}{HTML}{00aa5a}
\definecolor{W-NonStereo}{HTML}{aa9000}
\definecolor{ColorPerfo}{HTML}{283046}
\definecolor{ColorReduce}{HTML}{F9D510}
\definecolor{ColorNoPref}{HTML}{CB2727}

\newcommand{\stereotyped}{\textbf{\textsc{\fontfamily{phv}\fontsize{8.5}{1.2\baselineskip}\selectfont{\color{W-Stereo}stereo\color{M-Stereo}typed}}}\@\xspace}
\newcommand{\nonstereotyped}{\textbf{\textsc{\fontfamily{phv}\fontsize{8.5}{1.2\baselineskip}\selectfont{\color{W-NonStereo}non-stereo\color{M-NonStereo}typed}}}\@\xspace}

\newcommand{\MLstereotyped}{\textbf{\textsc{\fontfamily{phv}\fontsize{8.5}{1.2\baselineskip}\selectfont men-under\-paid$\times$\color{W-Stereo}stereo\-\color{M-Stereo}typed}}\@\xspace}

\newcommand{\WLstereotyped}{ \textbf{\textsc{\fontfamily{phv}\fontsize{8.5}{1.2\baselineskip}\selectfont women-under\-paid$\times$\color{W-Stereo}stereo\-\color{M-Stereo}\-typed}}\@\xspace}

\newcommand{\MLnonstereotyped}{\textbf{\textsc{\fontfamily{phv}\fontsize{8.5}{1.2\baselineskip}\selectfont men-under\-paid$\times$\color{W-NonStereo}non-stereo\color{M-NonStereo}typed}}\@\xspace}

\newcommand{\WLnonstereotyped}{\textbf{\textsc{\fontfamily{phv}\fontsize{8.5}{1.2\baselineskip}\selectfont women-underpaid$\times$\-\color{W-NonStereo}non-stereo\-\color{M-NonStereo}typed}}\@\xspace}

\newcommand{\WLfactor}{\textbf{\textsc{\fontfamily{phv}\fontsize{8.5}{1.2\baselineskip}\selectfont{women-underpaid}}}\@\xspace}

\newcommand{\MLfactor}{\textbf{\textsc{\fontfamily{phv}\fontsize{8.5}{1.2\baselineskip}\selectfont{men-underpaid}}}\@\xspace}

\newcommand{\Wlower}{\textbf{\textsc{\fontfamily{phv}\fontsize{8.5}{1.2\baselineskip}\selectfont{women-underpaid}}}\@\xspace}

\newcommand{\PGfactor}{\textbf{\textsc{\fontfamily{phv}\fontsize{8.5}{1.2\baselineskip}\selectfont{paygap-direction}}}\@\xspace}

\newcommand{\Colorfactor}{\textbf{\textsc{\fontfamily{phv}\fontsize{8.5}{1.2\baselineskip}\selectfont{color}}}\@\xspace}

\newcommand{\BlueUnderpaid}{\textbf{\textsc{\fontfamily{phv}\fontsize{8.5}{1.2\baselineskip}\selectfont{\color{M-Stereo}blue-underpaid}}}\@\xspace}
\newcommand{\PinkUnderpaid}{\textbf{\textsc{\fontfamily{phv}\fontsize{8.5}{1.2\baselineskip}\selectfont{\color{W-Stereo}pink-underpaid}}}\@\xspace}
\newcommand{\GreenUnderpaid}{\textbf{\textsc{\fontfamily{phv}\fontsize{8.5}{1.2\baselineskip}\selectfont{\color{M-NonStereo}green-underpaid}}}\@\xspace}
\newcommand{\YellowUnderpaid}{\textbf{\textsc{\fontfamily{phv}\fontsize{8.5}{1.2\baselineskip}\selectfont{\color{W-NonStereo}yellow-underpaid}}}\@\xspace}

\newcommand{\AFbox}[3]{
\begin{tcolorbox}[sharp corners,
enhanced,
attach boxed title to top left={xshift= \parindent, yshift=#1,yshifttext=0mm},
bottom= -0.5em,
  colback=white!95!gray,
  colframe=black!5!black,
  colbacktitle=white!95!gray,
  title=\textbf{#2}, 
  coltitle=black!100!black,
  boxrule=-1pt,
  toprule=1pt,
  boxed title style={size=small,colframe=white!95!gray},
]
  #3
\end{tcolorbox}
}
\newcommand{\motherbox}[1]{
\begin{tcolorbox}[
boxrule=0pt,
frame hidden,
sharp corners,
enhanced,
colback= white!95!gray,
borderline west={3pt}{0pt}{black!8},
left=3pt,
top =-0em,
bottom=0.5em,
before skip=1em,
after skip=1em
]
#1
\end{tcolorbox}
}

\definecolor{ReallyLightGray}{HTML}{fafafa}
\colorlet{shadecolor}{ReallyLightGray}
\usepackage{lipsum}
\newenvironment{shadedquotation}
 {\begin{shaded*}
  \quoting[leftmargin=2pt,rightmargin=2pt, vskip=0.5pt]
 }
 {\endquoting
 \end{shaded*}
}
\usepackage[most]{tcolorbox}

\usepackage{mathptmx}                  

\newlength{\myLength}	
\newlength{\mytextsize}

\usepackage{tikz}

\usepackage{wheelchart}

\newcommand{\minipie}[2]{
\hspace{-0.5em}
\raisebox{3pt}[0pt][0pt]{
\resizebox{\myLength}{!}{
\begin{tikzpicture}[baseline = 0\mytextsize]
\wheelchart[
  data sep=0,
  inner data =,
  pie,
  expand list=true,
  contour = black!50,
  slices style={
  \WCvarB,
  }
  ,
]{%
  #1/#2/,%
  100-#1/white/
}
 \end{tikzpicture}
}
}}

\begin{document}





\firstsection{Introduction}

\maketitle
When selecting colors to represent women (or girls) or men (or boys), one association likely comes to mind: \textcolor{W-Stereo}{pink} and \textcolor{M-Stereo}{blue}. This gender-color association is extremely present in news media, fashion, children's toys, and many other areas \cite{cunningham2011coloractivate, pomerleau1990pink, delgiudicePinkBlue2017}. Yet, choosing colors to represent 
different genders is a significant 
decision, as colors can convey messages, ideas, and stereotypes. Social cognition researchers have demonstrated that pink and blue colors are cues strong enough to trigger stereotyped behaviors that lead to biased decisions \cite{cunningham2011coloractivate}. Therefore, choosing pink and blue goes far beyond ``just picking colors'' as the use of these specific colors can alter the behavior of both children and adults.  

Despite the extensive research on colors in the visualization community---especially regarding efficacy, accessibility, and emotions---there is a significant gap in understanding how color-category associations holistically influence the reader of a data visualization. The most relevant studies evaluate the usefulness of associating colors with their semantics for visualization perception \cite{linresonant2013,schlossMappingColor2019} and show that semantically resonant colors help users read charts faster. 
However, when making decisions on social issues, it is better to aim for fair rather than quick decisions \cite{Cabric2023SocioDemographic}.
In this paper, we look at the use 
of color in visualization in a difficult, high-stakes decision influencing people. 
In particular, we investigate how gender-color associations affect visualization-driven decisions within a context characterized by gender biases: salary decisions. 
Our work is meant to illuminate possible connections between gender-color association and users' decision-making. In particular, our goal is to address the following research question: \textbf{To what extent can the use of gender stereotyped colors 
influence allocation decisions in a compensation 
scenario?}


We conducted two crowdsourced experiments with human resources (HR) experts.  
We evaluated how differently colored visualizations affected the way participants allocated limited funds in a fictitious company: they could opt to either reduce the gender pay gap in a fictitious company or reward employees for their performance. The two studies differ in the presence (Study 1 \cref{sec:expe1}) or absence (Study 2 \cref{sec:expe2}) of a legend in a data visualization indicating what color-gender association was used.
Our \Colorfactor conditions were \stereotyped: pink and blue and \nonstereotyped: yellow and green. We chose yellow and green because as prior work establishes, this is one of the least used color combinations to represent gender \cite{Cabric2024Gender}. In both studies, we varied which group was underpaid as the second study factor. In other words, the gender pay gap direction was either unfavorable to women (\WLfactor) or men (\MLfactor). Our main results highlight the complex nature of color-gender associations in visualization-driven decision-making tasks. Indeed, on their own, colors did not seem to influence decisions in any direction. Yet, results from Study 1 demonstrated that participants tended to favor a reduction of the gender pay gap when their own gender was underpaid and to reward performance more when the opposite gender was underpaid. Interestingly, in Study 2, when colors were the only attribute enabling gender differentiation, this effect disappeared.

Our work is one of the first to study the role of visualization in gender pay decisions. If we want to address 
the gender pay gap problem, we need to look at decision-makers and the tools they use as a basis for their decisions. We hope this paper will be one of many to study how stereotypes can affect the perception of demographics conveyed by data visualization and can pave the way for future studies seeking 
to reduce inequity through data visualization.

In summary, our paper is one of the first to investigate the impact of color in visualizations on decision-making. As such, it serves as a template for others to follow and highlights the many challenges of studying the impact of color on complex decision-making tasks. Our study results contribute to a deeper understanding of the effect of stereotyped colors in allocation decision-making and highlight the importance of individual characteristics (gender) and social norms in visualization-supported decision-making regarding pay. Finally, the paper highlights a number of interesting research avenues for the visualization community aiming to support decision-making in high-stakes domains.
\section{Related Work}

Our work is situated at the intersection of research in colors, decision-making biases, and non-neutrality of data visualizations.


\subsection{Color and gender}
Colors and their associated meanings have been studied across various fields, including vision science, psychology, and data visualization. In data visualization, researchers have mainly focused on how colors can affect task performance \cite{linresonant2013,wangOptimizingColor2019,tsengMeasuringCategorical2023,smartMeasuringSeparability2019}. One effective strategy to enhance task performance is to align color choices to categories with existing color-concept associations. An example of such association is representing the category ``banana'' in a bar chart using a yellow bar. Lin and colleagues \cite{linresonant2013} found that such semantically resonant colors facilitated chart reading and category interpretation. This study, among others \cite{wangUnlockingSemantic2024,schlossSemanticDiscriminability2021,schlossMappingColor2019,elassadySemanticColor2022}, emphasizes the significant role of semantic meaning in color choices. 
Recently, Karen Schloss developed the \textit{Color Inference Framework} \cite{schlossColorSemantics2024}, which models how people use color-concept associations to make judgments about the world.

When it comes to gender visualizations, the association of ``pink for women and blue for men'' is widespread.  While there is a debate about the origin of this association, its popularization began in the mid-twentieth century. Nowadays, the use of pink and blue (or similar colors such as red and blue \cite{Cabric2024Gender}) to represent women and men is highly prevalent\cite{cunningham2011coloractivate, delgiudicePinkBlue2017,pomerleau1990pink}. 
Consequently, numerous researchers have investigated this gender-color association, resulting in studies on the effect of pink or blue clothes \cite{ishiiEffectPink2019}, the impact of pink in marketing for women \cite{atkinsonJustColour2024}, and brand perception \cite{hess2016coloractivate}. From a cognitive standpoint, researchers have also illustrated how the use of color influences the behavior of both adults and children \cite{cunningham2011coloractivate}. In a series of six studies, Cunningham and Macrae \cite{cunningham2011coloractivate} found that stereotypical colors alone could alter behaviors in children and adults. Children selected stereotypically colored furniture for girl or boy bedrooms. For adult participants, 
their studies showed that gender-typed colors triggered the activation of associated categorical knowledge and made participants slower when the stereotypical color did not match the actual gender association of an object (such as a first name, the picture of a face, or a typically male or female object such as a lipstick or a hammer).  

In this paper we contribute to this literature by 
investigating whether a stereotyped gender-color association 
influences a visualization-driven compensation decision. We limit our scope to 
the gender-color association. However, we acknowledge 
that colors have many other properties and/or associations, such as their emotional resonance (see \cite{jonauskaiteWeFeel2025} for a review), that represent an interesting avenue for future study.


\subsection{Decision-making, stereotypes and biases}

Biases are unconscious drifts in how individuals interpret information or make decisions. They can be beneficial when individuals are faced with a vast number of alternatives and have to make a fast decision. Yet they are detrimental when judgment is swayed by uncontrollable human factors, such as prior beliefs or past experiences, which lead to unfair decisions. Ideally, and to reduce 
inequity, decision-makers should be basing their decisions based on objective 
data interpreted without personal bias. 

One of many triggers of biases is the perception of stereotyped content \cite{Bodenhausen2005Stereotypes,Bodenhausen1988stereojuror,Koch2015MetaGender}. For a stereotype to affect cognition, it must first be activated and then applied \cite{kunda2003stereoactivation}. Exposure to stereotyped content can influence various stages of the decision-making process, from categorization to interpretation and response \cite{Bodenhausen1998StereotypeInhibition}. At each stage, the presence of stereotyped content can support or inhibit the application of the stereotype. In a meta-analysis, Koch and colleagues \cite{Koch2015MetaGender} examined the relationship between gender stereotypes and biases in organizational (mostly hiring-related) decision-making, finding that stereotypes can significantly influence judgments. Among other results, the authors found that male decision-makers tended to favor men and that a motivation to make careful decisions led to less biased decisions.

Cognitive biases also play a crucial role in how individuals interpret and respond to data visualizations. These biases can mislead individuals in their interpretation of data and impact decisions made using data visualizations
\cite{dimara2019attraction,Dimara202TaskBased,Xiong2023BeliefsInfluence}. Xiong and colleagues \cite{Xiong2023BeliefsInfluence} showed that prior beliefs can influence how individuals correlate two attributes, as the participants interpreted the data following their existing beliefs.

Despite increasing regulatory and societal pressure to address the gender pay gap---which is the phenomenon that women are paid less than men despite having equivalent qualifications and job responsibilities \cite{anderson2019firm}---there is very limited research, if any, on the role of visualization, or decision making support tools more generally, in compensation decisions. While the gender pay gap has been a focal point of scholarly investigation for decades (\eg Blau \& Kahn \cite{blau2017gender}), only in recent years have scholars started to focus on decision-making in this context, and then they typically focus on how to optimally allocate raises in order to close unexplained gender pay gaps \cite{anderson2019firm,anderson2023bridging}.

To our knowledge, our study is the first to investigate the extent to which the activation of a stereotype through data visualization can influence 
pay decisions. When a visualization includes gender data, stereotype biases may 
harm the individuals represented in the data \cite{Schwabish2021Harm}. Consequently, our research aligns with existing literature highlighting the potential harmfulness and non-neutral nature of sociodemographic data visualization and contributes to the literature on compensation decision-making in the presence of unexplained gender differences.

\subsection{Non-neutrality of sociodemographic data visualization}

For a long time, researchers and readers have viewed data visualization as neutral, objective, and devoid of societal implications. But recently, many researchers have outlined how data visualization are not necessarily neutral and can cause harm \cite{dhawka2025SocialConstructionVisualization,Cabric2023SocioDemographic,Schwabish2021Harm,correllEthicalDimensions2019,dorkCriticalInfoVis2013,BaumerItsPolitical2020,holderDispersionVs2022}. In a forward-looking article, D{\"o}rk and colleagues provided a critical framework for visualization that promotes disclosure, plurality, contingency, and empowerment \cite{dorkCriticalInfoVis2013}. For instance, when visualizing gender data, it is essential to consider how gender non-conforming people can be included in visualizations without discriminating or overexposing them, especially when their numbers are very low \cite{Cabric2023SocioDemographic}.  By designing and then examining visualizations through one or more of the aspects proposed by D{\"o}rk and colleagues' framework, we can better identify the potentially harmful effects of visualization design. 

In their ``Do no harm guide,'' Schwabish and colleagues emphasize that when data about people is represented,  careful design is essential to avoid misrepresentation and thus, harming those people \cite{Schwabish2021Harm}. Beyond the harm that can arise from misrepresenting or failing to represent certain social categories, visual choices can also negatively affect the lives of individuals represented in decision-support visualizations \cite{dhawka2025SocialConstructionVisualization}. Holder and Xiong \cite{holderDispersionVs2022} showed how different visualization designs (bar charts, jitter plots, or prediction intervals) lead to misinterpretations of inter-group disparities, therefore reinforcing stereotypes. In this work, we aim to identify how 
data visualization with 
stereotypical gender colors 
may impact pay decisions for 
the represented gender groups.

\section{Experimental Design}\label{sec:expedesign}
To evaluate whether stereotyped colors affect decision-making, we conducted two crowdsourced user experiments. We designed 
a task 
focused on salary adjustments, 
a context in which gender bias is well documented and leads to persistent gender pay gaps. Each experiment focused on 
the same two factors: 
\begin{description}[\setlabelphantom{\Colorfactor:}\compact]
\item[\Colorfactor:] stereotyped or not
\item[\PGfactor:] which of the two employee groups was underpaid. Groups were differentiated by color and gender legend in Study 1 and only by color in Study 2.
\end{description}
Each study had a 2 (\Colorfactor) x 2 (\PGfactor) between-subjects design. The only difference between the experiments was the presence (Study 1 in \cref{sec:expe1}) or the absence (Study 2 \cref{sec:expe2}) of a legend that linked the colors to the gender of the represented employees.

In this section, we detail the scenario, stimuli, and tasks presented to participants. Additionally, we present the data collected and the analysis performed. The two experiments were inspired by Verma and colleagues' \cite{verma2023FairAllocation} studies on data visualization to promote fair allocation of resources. Our two experiments received ethical approval from our institutional ethics committee. A video and the source code for the study are available in the supplemental material (see \cref{sec:supplemental_materials}).

\begin{figure*}[tb]
 \includegraphics[width=\linewidth]{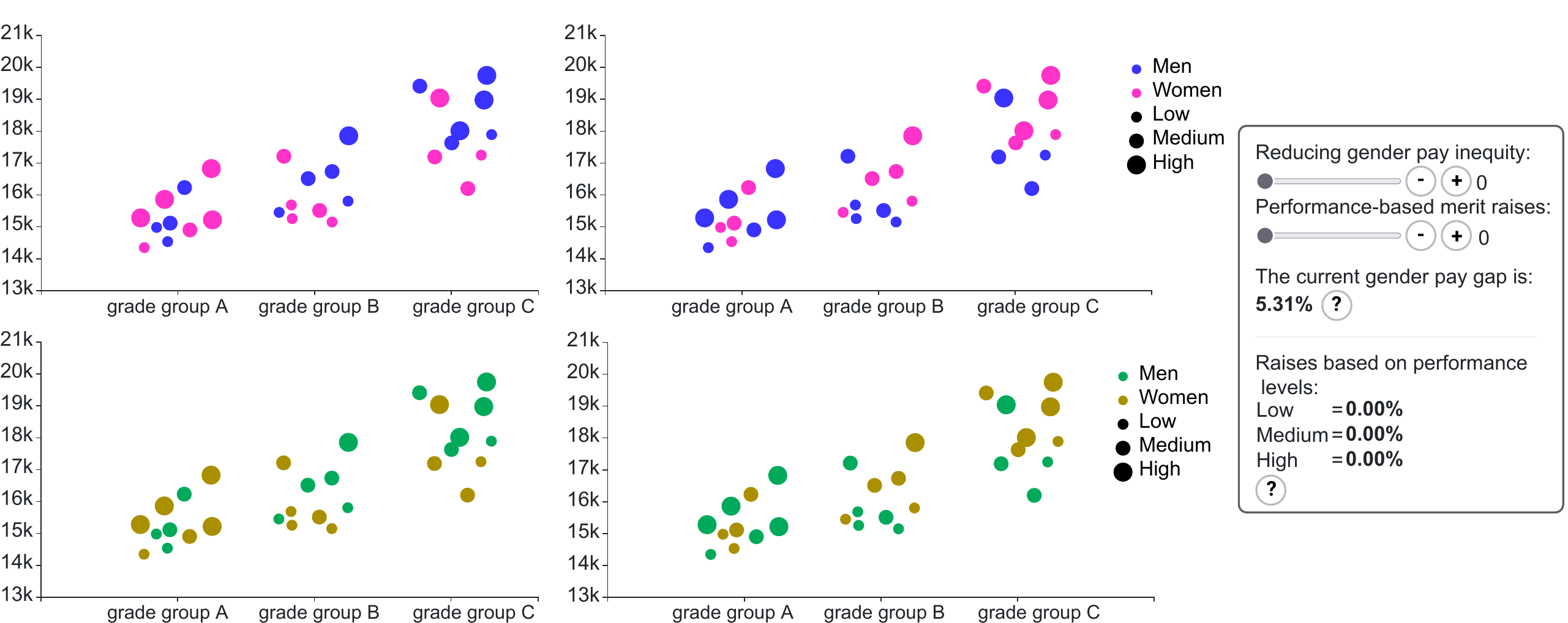} 
 \caption{The interactive panel, the legend used in Study 1, and the scatterplots used as stimuli for the allocation task (from top to bottom and left to right): \WLstereotyped, \MLstereotyped, \WLnonstereotyped and \MLstereotyped. In the interactive panel, the sliders are positioned at the top and the two pieces of information, actual pay gap and performance rewards, are below. Help buttons were available to display more information on-demand. In Study 2, we re (moved the color legend (\textcolor{M-Stereo}{\CIRCLE}/\textcolor{M-NonStereo}{\CIRCLE} Men, \textcolor{W-Stereo}{\CIRCLE}/\textcolor{W-NonStereo}{\CIRCLE}Women).}
 \label{fig:scatterplots}

\vspace*{-2em}
\end{figure*}

\subsection{Scenario}
Decision-making in the context of pay adjustments is prone to stereotyped behavior \cite{Chaxel2015Stereotypes,Bodenhausen1985stereoheuristics}. In contrast to perceptual tasks, decision-making involves different cognitive steps in which the activation of stereotypes can play a role and bias the outcome (\ie the decision) \cite{Bodenhausen1998StereotypeInhibition}. 

Designing a decision-making task requires contextual information to engage participants in careful decision-making. To ensure the possibility of a stereotype activation, we focused on salary adjustments within a company. In the popular media, the gender pay gap often discussed is the absolute pay gap, that is, the difference in the average pay of men and women. In contrast, the pay gap considered in our scenarios is the \textit{adjusted pay gap}, the difference in pay after accounting for factors such as performance, experience, and seniority. In our studies, participants were shown a visualization of factors that the adjusted pay gap corrects for, in our case grade group (the complexity of a job) and performance. For our task, participants received the following prompt: 

\vspace{-1em}
\begin{shadedquotation}
You are the person in charge of your company's annual salary review. The CEO has allocated 10,000 StudyCoin to increase employees' salaries — your colleagues. The CEO instructed you to allocate the 10,000 between two objectives: 
    \begin{itemize}
    \item reducing pay inequity focusing on gender differences
    \item allocating merit raises based on last year’s individual performance.
    \end{itemize}
    
\end{shadedquotation}
\vspace{-1.5em}
\subsection{Stimuli and data}\label{sec:designStimuliData}

To examine the impact of stereotypical colors on visualization-supported decision-making, we created one scatterplot per condition (see \autoref{fig:scatterplots}). Each displayed data for 30 employees, comprising 15 women and 15 men, along with their annual salaries. Gender was represented using either \stereotyped colors (pink for women, blue for men) or \nonstereotyped colors (yellow for women, green for men), our \Colorfactor factor. 
Each represented employee was categorized into one performance level\footnote{The participants received the information that the performance evaluation was robust and unbiased.} from 1 to 3, represented by size of the dots on the scatterplot. Additionally, employees were divided into three grade groups
(A, B, and C) positioned along the x-axis. In grade group A (and similarly for B and C), the average annual salary was approximately 15,300 StudyCoin (for A, from 14,300--16,800), 16,100 (for B, from 15,100--17,800) and 18,100 (for C, from 16,200--20,000). The salaries were generated to reflect an adjusted gender pay gap of approximately 5\%, which is higher than the adjusted pay gap in some countries, such as Australia, France or Canada, \cite{Glassdoor2019PayGap} and which is the upper limit of the new EU regulation\cite{EU2023PayGapDirective
}. To design the two conditions of the \PGfactor factor, we first created the \WLfactor visualizations and then swapped each dot's gender code to create the \MLfactor conditions (\ie women became men and vice-versa), which ultimately appears as color swapping. To enhance the realism of our decision-making task, the dataset was created by one author who specializes in modeling gender salary data.

To ensure that participants understood the data, we provided supplementary information next to the visualization, including the current adjusted gender pay gap and the percent allocation for each performance group (see \autoref{fig:scatterplots}). To simplify the decision-making task, all employees in a given performance group received the same percentage of their base salary. The allocation task focused on two sliders: the top one allocated funds to reduce gender pay inequity, and the bottom one allocated funds to reward performance.

The visualization was interactive. As the participant moved the sliders, both the visualization and the supplementary information were updated in real-time. This interactive feature was meant to help participants understand the outcomes of their decisions and explore various decision possibilities. In addition, when participants started to modify the current allocation (\ie when their cursor hovered over the sliders), a light gray vertical line showed the maximum salary that an employee could get from the decision (see the video in the supplemental material).

\subsection{Task}\label{sec:designTask}
We designed the task to fit an ecological scenario where a company wants to distribute year-end bonuses based on different criteria; it is typical that companies combine their equity correction and merit adjustments in a single salary update. The decision-maker had a total budget of 10,000 StudyCoin to allocate. The participants decided on the amount of StudyCoin allocated between two options: to reduce the adjusted gender pay gap or to reward performance. Importantly, this was not a binary decision; rather, the participants could decide the amount allocated to each objective, enabling participants to make more nuanced decisions. 
Although rewarding performance is generally perceived as a fair practice \cite{hartmannPerceivedFairness2012}, it is still susceptible to biases \cite{neschenGenderBias2021}. We expected these two alternatives to provide a meaningful tradeoff, as both objectives could be perceived as equitable, leading to competing goals \cite{Frisch2001Tradeoff}. We further considered the use of these competing goals as a way to identify the appearance of stereotyped behavior (\ie a difference in the allocation in the \Colorfactor conditions). 
 
Participants completed their resource allocation with two linked sliders both starting at 0. Adjusting one slider did not impact the other unless the total budget had been reached. In such cases, participants had to decrease one slider to make StudyCoins available to be allocated to the other goal. Using two sliders enabled us to start the decision-making task with no initial decision (both sliders were at 0). In an alternative design with just one slider, we would have needed to choose its starting position--- probably in the middle ---which is already a possible solution to the allocation task and which would potentially anchor the participants’ decisions. 

\subsection{Procedure}

We published the two studies online on the Prolific crowdsourcing platform. The study consisted of five steps: the consent form, an explanation of the scatterplot, information and explanation of the interactive features, the task, and a series of post-task questions.

\paragraph{Scatterplot explanation.} After reading the study information on Prolific, participants were redirected to our website hosting the study. Each participant was assigned a condition upon entering, and once they consented to participate, they could start the study. The first phase was designed to verify that participants understood how to read our visualization. Participants learned how to interpret the y-axis (salaries), the x-axis (grade group), and size (performance). In addition, participants were introduced to the concept of jitter, which we used to move employee dots to avoid overlap. At the end of each explanation, participants had to correctly answer a question about what they had just learned to move on to the next step. At the end of the explanation stage, participants took a comprehension test with three single-choice questions. They had two attempts to pass. If they failed, participants were sent back to Prolific, and they did not become a part of the study (and they were not paid). After this test, we briefly introduced the colors used in their assigned condition (\ie either \stereotyped or \nonstereotyped). In Study 1, a color legend was introduced to users, and it remained visible throughout the following steps.

\paragraph{Information and explanation for interactive features.} This phase began with presenting the task to participants (see \cref{sec:designTask}). During the explanation, participants were asked to experiment with an amount different from the task to avoid priming their final decision. We introduced the sliders used for the allocation task. To proceed, participants had to successfully move the two sliders to three locations to allocate 1) all funds towards reducing gender pay inequity, 2) all funds to reward performance, and 3) half to each goal. After the participant completed these tasks, we introduced the interactive update of employees' salaries. To ensure participants understood that moving the sliders triggered a visual update, they needed to interact with the sliders for at least 10 seconds before proceeding. The final step of this phase was to introduce the additional information provided in the visualization, the gender pay gap and the allocation to performance (see \autoref{fig:scatterplots}). At this point in the study, participants could interact with the sliders for as long as they desired to observe all the interactive features in action.

\paragraph{The task.} Before performing the allocation task, we asked participants to describe the salary distribution in a few words using a text field. During this task, they were unable to move the sliders or view the information about the pay gap.  Once they submitted their description, they performed the allocation task detailed in \cref{sec:designTask}. The participants could not submit their decision until at least 30 seconds had elapsed. After this timeframe and once participants allocated the entire 10,000 StudyCoin, they could click a button to submit their decision. Before submitting their decision, a pop-up informed participants that there would be only one allocation task and that reconsidering the decision after the submission would not be possible. At this point they could still go back and re-consider their allocation. Similar to other resource allocation studies \cite{verma2023FairAllocation}, participants performed a single allocation task to prevent them from correcting their strategies after giving their explanations. We were concerned that allowing them to 
reflect on the criteria they used to make the decision, would prime them for future tasks. Having a single task prevents participants from blindly repeating their approach across multiple trials even under changed parameters.

\paragraph{Post-task questionnaire.} Participants were asked to explain their decision, express their satisfaction, and indicate whether they are decision-makers in their daily job. Only the first question about decision explanation was mandatory and required at least a 100-character response. On the question page, the result of their decision (visualization and additional information) was displayed at the top.

Once the answers were submitted, participants were required to complete a short form of the Ambivalent Sexism Inventory (ASI) \cite{rollero2014PsychometricProperties}. The ASI questionnaire allowed us to capture the degree of both ambivalent and hostile sexism of the participants\cite{glick1996AmbivalentSexism}, which we considered to be a possible contributing factor to their decisions. Participants had to answer 12 questions on a 6-point Likert scale. Completing the questionnaire marked the end of the study, and participants were asked to confirm their submission to Prolific. Upon successful completion, participants received a \pounds3.46 compensation in Study 1 and received \pounds3.97 compensation in Study 2, reflecting minimum wage for approximated task times of 21 and 24 minutes, respectively. 

\subsection{Measurements}\label{sec:designMeasurement}
The primary measurement in our studies was the allocation between the two options (rewarding performance or reducing the pay gap). A participant who allocated 100\% of the 10,000 StudyCoin to reducing the gender pay gap would be given a \textit{decisionValue} of 100. A participant who allocated 100\% to rewarding performance would be assigned a \textit{decisionValue} of -100. A participant who used a 50/50 strategy would have a \textit{decisionValue} of 0. We categorized the strategies into three groups based on the allocations made: those who primarily rewarded performance (\textit{decisionValue} < 0), those who primarily reduced gender inequity (\textit{decisionValue} > 0), or those who showed no preference for either strategy ("No Preference" strategy,  \textit{decisionValue} = 0).

To compute the ASI results, we divided the questionnaire into two scales (benevolent sexism and hostile sexism) and averaged the scores of each scale, as suggested by the authors \cite{glick1996AmbivalentSexism,rollero2014PsychometricProperties}. Since we saw no evidence of an association between scale results and participants' decisions, we do not discuss these further. 

\subsection{Data analysis}\label{sec:designDataAnalysis}

We analyzed participants' allocation decisions using two different methods. First, we estimated the average  \textit{decisionValue}, with summary statistics (\ie average) and visualization of  95\% confidence intervals,  inferring our insights from point and interval estimates (similar to \cite{verma2023FairAllocation}). Second, we assigned each participant's allocation to one of three strategies (see \autoref{sec:designMeasurement}). 
Both methods complementarily contribute to the inferences we draw about our results. We broke down the results by study condition and by participants' gender. As participant gender was not a planned factor in our studies, the distribution of gender within each condition was not equal, which limited our ability to make statistical comparisons. Therefore, we analyzed the conditions by participant's gender using the assigned strategy only. 

In our studies, three participants identified as non-binary. Consequently, a detailed and comparative analysis of their results was not possible. However, at the end of the gender analysis, we provide their results for comparison. We eliminated two participants who did not declare their gender from the study of gender differences.

The first author coded the description of the visualization that took place before the allocation task and the free-text answers of the post-task questionnaire for analysis. The coding themes were discussed and validated during meetings with the other authors. The codes were designed to highlight specific strategies used by participants, opinions on the pay gap (both the desired outcome and what is considered ``fair''), perceptions of task complexity, and overall engagement in the study. For the second study, where we removed the color legend, we coded when participants associated colors with represented genders and in what condition they made these assumptions.

\subsection{Population sample}
Given the focus of our study, we recruited participants who worked in HR. Screening criteria included being of legal age and not being color-blind. For both studies, we targeted a sample size of 280 participants, with 70 people per condition. This target would give us .95 power to detect a “medium” Cohen’s d effect size of .25 between any two factors, as computed by the G*Power software
for differences between independent means. Participants of Study 1 and Study 2 were different.

\section{Study 1: Gender-Color Association + Legend}\label{sec:expe1}
Our first study investigated the potential effect of \stereotyped colors following the experimental design detailed in \cref{sec:expedesign}, where a legend associated color with gender. Based on our literature analysis of gender stereotyping, we developed the following hypotheses: 
\begin{itemize}
    \item \textbf{HP1}: Participants in the \stereotyped gender-color condition will show different allocation decisions compared to those in the \nonstereotyped condition.
    \item \textbf{HP2}: Our participants were domain experts aware of the gender pay gap and that women are more typically underpaid; therefore, they will prefer reducing gender inequity in the \Wlower conditions compared to the \MLfactor ones.
\end{itemize}

\begin{table}
    \centering
        \caption{Study 1 gender breakdown}
        \vspace{-0.5em}
   \small
    \begin{tabular}{p{0.35\columnwidth}p{0.40\columnwidth}p{0.06\columnwidth}}
    \hline
         &  Women including trans women & 40\\
         &  Men including trans men  & 29 \\
         & Non-binary & 0 \\
        \multirow{-4}{0.35\columnwidth}{\MLnonstereotyped } & Rather not to say & 0 \\
        \midrule
         &  Women including trans women& 40\\
         &  Men including trans men & 28 \\
         & Non-binary & 1 \\
        \multirow{-4}{0.35\columnwidth}{\WLnonstereotyped} & Rather not to say & 0 \\
        \midrule
        &  Women including trans women& 35\\
         &  Men including trans men & 33 \\
         & Non-binary & 1 \\
        \multirow{-4}{0.35\columnwidth}{\MLstereotyped} & Rather not to say & 0 \\
        \midrule
          &  Women including trans  women& 38\\
         &  Men including trans men & 31 \\
         & Non-binary & 0 \\
        \multirow{-4}{0.35\columnwidth}{\WLstereotyped} & Rather not to say & 0 \\
         \hline
    \end{tabular}
    \label{tab:demo_study1}
    
\vspace{-2em}
\end{table}

\subsection{Participants}
Out of 280 enrolled participants, four experienced data logging problems. The remaining 276 participants were evenly distributed across the four conditions (69 each).  The gender distribution of participants across these conditions is detailed in  \autoref{tab:demo_study1}. It is important to note that we did not collect data from individuals who dropped out of the study after consenting, making attrition analysis infeasible.



\subsection{Analyses and results}

Overall, more participants prioritized reducing the gender pay gap (136\minipie{49.2}{ColorReduce}{\small 49.2\%} ) rather than rewarding performance (110\minipie{39.9}{ColorPerfo}{\small 39.9\%}) or expressing no preference (30\minipie{10.9}{ColorNoPref}{\small 10.9\%}) (see \cref{fig:countgeneral}). Participants who chose to primarily reduce the gender pay gap allocated more money on average for this cause (on average 34.7, CI[30.3; 39.16]) compared to those who preferred rewarding performance (on average 36, CI[31; 41] of the absolute value, see \cref{sec:supplemental_materials} Supplemental Material).

When we divide up the analysis by \PGfactor and \Colorfactor, we see in \autoref{fig:cis} and \autoref{fig:count_conditions} that the two \Wlower\ conditions led to strategies that favored a reduction of the gender pay gap. \WLstereotyped and \WLnonstereotyped resulted in a mean of 12.6 (CI[2.0; 23.3]) and 13.1 (CI[3.8; 22.5]) while \MLstereotyped and \MLnonstereotyped resulted in a mean of -2.8 (CI[-14.2; 8.8]) and 1.2 (CI[-8.3; 10.8]).
The analysis of the allocation for the \PGfactor and their 95\% confidence intervals reflects this allocation: the mean for \WLfactor equaled 
12.9 CI[5.9; 20] and for \MLfactor it equaled -0.8 CI[-8.3; 6.6]), thereby providing evidence in favor of \textbf{HP2} (see \autoref{fig:cis} middle). Conversely, the colors used to represent gender did \emph{not} influence allocation decisions, neither the 
strategies nor the average allocation amount, and therefore we do not observe any evidence for \textbf{HP1} (see \cref{fig:cis}).


\subsection{Analysis by gender}\label{subsec:study1:ana-gender}



\begin{figure*}[!t]

 \captionsetup[sub]{singlelinecheck = false, format= hang, justification=raggedright, font=footnotesize, labelsep=space}
  \centering
     \begin{subfigure}[b]{\linewidth}
         \centering
         \includegraphics[width=0.9\linewidth]{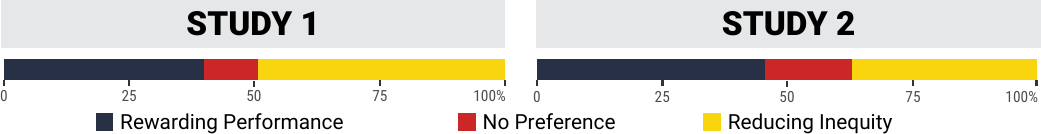}
         \caption{Types of decision, in percent, normalized to the number of participants in Study 1 (left) and Study 2 (right).}
         \label{fig:countgeneral}
         
      \vspace*{0.4em}
     \end{subfigure}
    \begin{subfigure}[b]{\linewidth}
         \centering
         \includegraphics[width=0.9\linewidth]{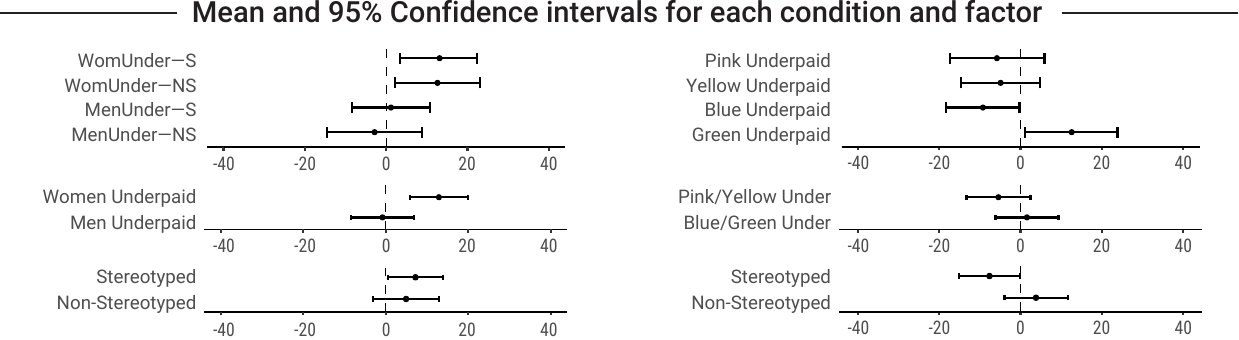}
    \caption{Average of the \textit{decisionValue} and the 95\% confidence intervals for all the conditions (top), the \PGfactor (middle) and the \Colorfactor~(bottom) in Study 1 (left) and Study 2 (right).}
         \label{fig:cis}
         
      \vspace*{0.1em}
     \end{subfigure}
     \begin{subfigure}[b]{\linewidth}
         \centering
         \includegraphics[width=0.9\linewidth]{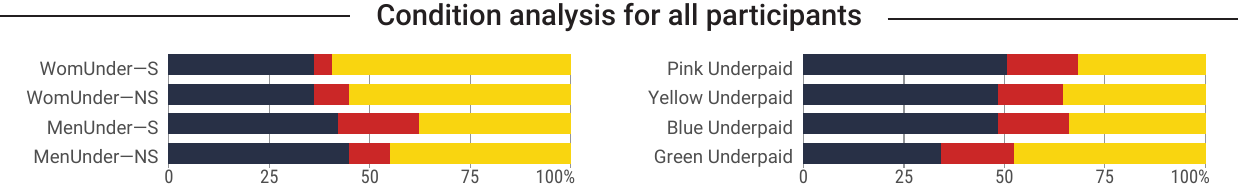}
    \caption{Types of decision, in percent, normalized to the number of participants in Study 1 (left) and Study 2 (right) for each condition.}
         \label{fig:count_conditions}
         
      \vspace*{0.4em}
     \end{subfigure}
     \begin{subfigure}[b]{\linewidth}
         \centering
         \includegraphics[width=0.9\linewidth]{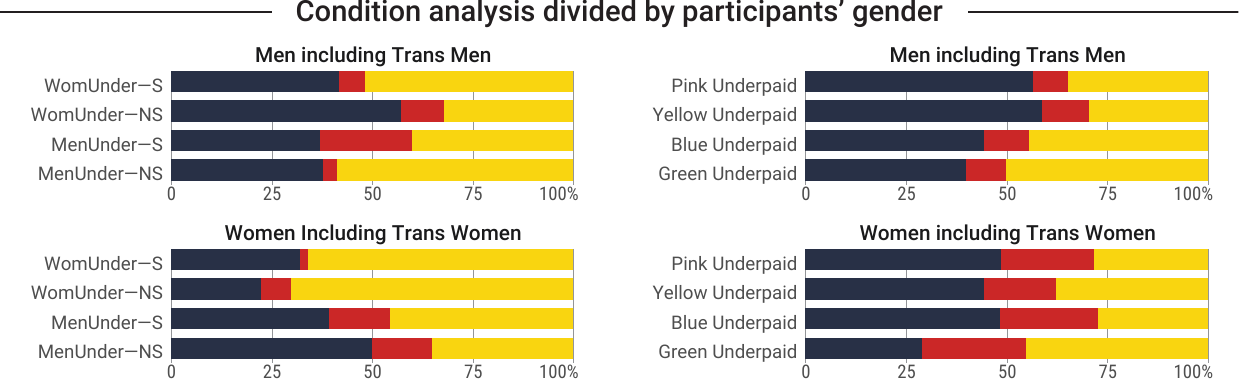}
    \caption{Types of decision, in percent, normalized to the number of participants in Study 1 (left) and Study 2 (right), divided by gender (facet) for each condition.}
         \label{fig:count_gender_conditions}
         
      \vspace*{0.4em}
     \end{subfigure}

     \begin{subfigure}[b]{\linewidth}
              \centering

         \includegraphics[width=0.9\linewidth]{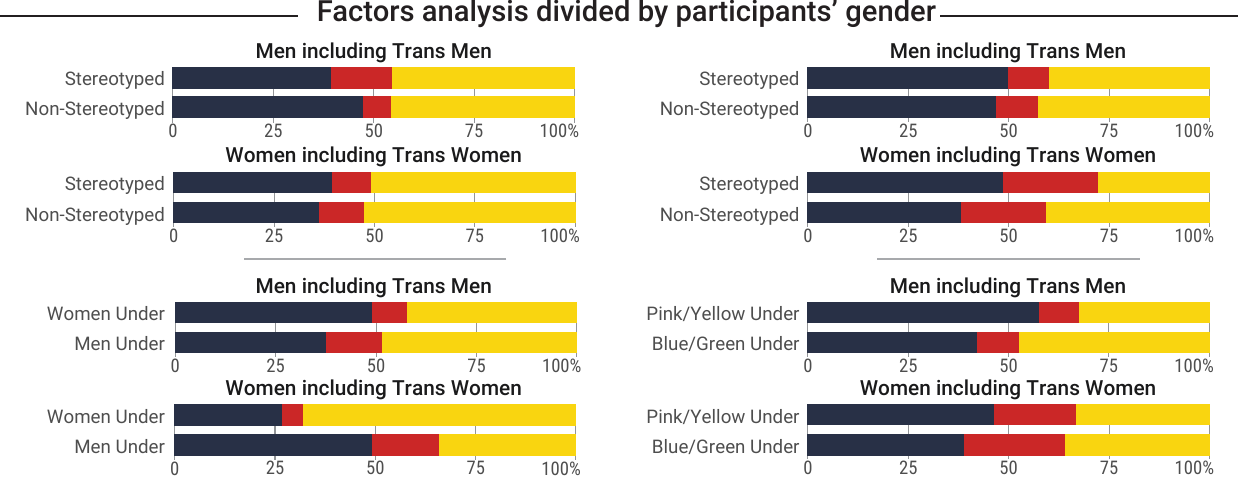}
         \caption{Types of decision, in percent, normalized to the number of participants in Study 1 (left) and Study 2 (right), divided by gender (facet) for the \Colorfactor (top) and \PGfactor (bottom)}
         \label{fig:count_gender_factors}
     \end{subfigure}
 \vspace*{-2em}

\end{figure*}

We broke down conditions by participants' gender in order to examine the role of this variable  (see \cref{fig:count_gender_conditions}). Participants were more likely to use strategies favoring inequality reduction when the underpaid group corresponded to their own gender (\ie their \emph{in-group} \cite{Tajfel1986SocialIdentity}). This effect is salient in the two \WLfactor\ conditions. When we only consider women individuals, 70\%\minipie{70}{black} in the \nonstereotyped and 65.8\% \minipie{65.8}{black} in the \stereotyped conditions chose to favor inequality reduction compared to 32.1\%\minipie{32.1}{black} and 51.6\%\minipie{51.6}{black} of men in the same condition. To a lesser extent, we see this effect in the \MLnonstereotyped condition where 58.6\%\minipie{58.6}{black} of men allocated more to reduce inequity compared to 35\%\minipie{35}{black} of women in this condition. As a result, participants were more likely to reward performance when the underpaid group was from the opposite gender. 

The analysis of the interaction between gender and the two factors separately confirmed the in-group effect described above (see \cref{fig:count_gender_factors}). However, our results showed no evidence that the interaction between colors and participants' gender affected allocation, as shown in \cref{fig:count_gender_factors}.  

Finally, the two non-binary participants allocated 68 and 74 for reducing the gender pay gap in, respectively, \MLstereotyped and \WLnonstereotyped conditions.

 

\captionsetup[sub]{font=normalsize,labelfont={bf,sf}}

\vspace{-0em}
\subsection{Qualitative analysis of open feedback}

After their allocation task, participants were asked to 1) explain their decisions in at least 100 characters, 2)  express if they were satisfied with their allocation and the resulting decision, and 3) indicate if they were compensation decision-makers in their jobs. These three answers added to the visualization description before the task constitute the analyzed feedback (see \cref{sec:designDataAnalysis}). On average, participants used 178 characters to explain their decision. Over 65\% of participants stated that they were being satisfied with their allocation decision. Only six participants did not answer all four questions. Overall, these results indicate good engagement of participants with the study. 

In their feedback, numerous participants mentioned the complexity of closing the gender pay gap with a single allocation, justifying their choice not to allocate 100\% of resources to reduce the pay gap. 
\motherbox{
\AFbox{-2.4mm}{Multi-year objective}{``Ideally it would be a multiple year plan to close the gender gap over time'' (P221)}}






Some participants mentioned having specific targets for the gender pay gap. These pay gap targets varied from 1\% (which would correspond to allocating the entire StudyCoin allowance to reduce it) to 4\%. In a few cases, participants argued that a gender pay gap of 5\% was already considered low, and this motivated their decision to prioritize a performance-based allocation.
\motherbox{
\AFbox{-2.4mm}{Opinion about pay gap level}{``Gender pay gap was already very low.'' (P249)}}

The idea of setting an objective was also mentioned by some participants who reported having a two-step strategy. Most of these participants decided to first reach a ``low enough'' gender pay gap --- without always specifying a target percentage --- and then allocate the remaining resources toward performance-based rewards. Conversely, few of them prioritized performance first, stating that rewarding performance is crucial to motivate low-performers and avoid discouraging high-performers.

\motherbox{
\AFbox{-2.4mm}{Two-step strategies}{``The pay gap isn't based on any skill so it should be solvable first. Then the merit aspect can be taken into account.'' (P44)}}


Some participants also mentioned the complexity of making such a decision with a limited amount of resources. Most of them perceived both alternatives as fair and thereby identified a higher level of resources as a potential solution to this tradeoff. 

\motherbox{
\AFbox{-2.4mm}{Need for more budget}{``I would have preferred fixing the gender gap completely but this was not possible with the allocated money'' (P151)}}



Additionally, as noted in \cref{subsec:study1:ana-gender}, some participants felt that their perceptions of the gender pay gap varied depending on whether women or men were underpaid, and they highlighted that it was unusual to see men underpaid. Therefore, participants who minimized the gender pay gap when men were underpaid favored rewarding performance.

\motherbox{
\AFbox{-2.4mm}{Direction of gender pay gap matters}{``also since women statistically earn less [in general] and here it doesn't seem to be a problem then I did not give it the same importance as to performance.'' (P196)}}

Finally, the notion of fairness was very present in the collected comments regardless of the decision made. This result highlights that perceived fairness is a very important driver in the decision strategy employed. However, fairness is a very subjective topic.

\motherbox{
\AFbox{-2.4mm}{Fairness}{``I distributed the allocation to what I perceived to be fair. '' (P7, who distributed 10,000 to reduce inequity)

``I [...] think that performance should be rewarded first and foremost because it is equal and fair.'' (P264, who allocated 10,000 for rewarding performance).}}



\subsection{Discussion and summary}

We initially designed this study to detect the effect of stereotyped gender-color association on visualization-driven resource allocation decisions.
However, we found no evidence of a color effect that would support our first hypothesis (\textbf{HP1}). Interestingly, we observed that participants, particularly women, were more inclined to reduce the gender pay gap when the underpaid group shared their gender. 

A possible limitation of our study, however, is that the color legend provided information about which color was associated with which gender in the \nonstereotyped condition. This was done to create a more realistic study condition. After all, visualizations typically contain a legend that describes the categories associated with colors. Yet, it is plausible that the presence of this legend overrode any effect of the color coding. This was our motivation for 
the second study, which used the same procedure but without the color legend.
\section{Study 2: Gender-Color Association, No Legend}\label{sec:expe2}
In the second study we replicated the initial study to investigate whether the presence of a legend may have overridden the effect of gender-color association, and we removed any legends linking colors to gender categories. As a result, participants had to make decisions independently of explicit gender categorizations. Yet, participants could infer the underpaid gender category based on a \stereotyped color association. We expected these modifications to alter the in-group effect observed in the previous study, especially for the \nonstereotyped conditions.


In the \nonstereotyped conditions, participants could not be sure what gender category was underpaid. 
Yet, given that individuals often interpret data through the lens of prior knowledge and beliefs \cite{Xiong2023BeliefsInfluence}, it is plausible that participants may systematically have associated the color representing the underpaid group with the women category.  Since the \nonstereotyped condition did not clearly show a pay gap direction, we focus our analysis of the \PGfactor here on the color (yellow or green) that represented the underpaid group. Therefore, in this study, we refer to our four conditions as \BlueUnderpaid and \PinkUnderpaid for the \stereotyped conditions and \GreenUnderpaid and \YellowUnderpaid for the \nonstereotyped ones.
We propose two new hypotheses: 
\begin{itemize}
    \item \textbf{HP3}: Participants will allocate the same amount of money regardless of \PGfactor in the \nonstereotyped condition. 
    \item \textbf{HP4}: Women participants will allocate more money to reduce the gender pay gap in the \PinkUnderpaid condition than in the two \nonstereotyped conditions.
\end{itemize}



\begin{table}
    \centering
        \caption{Study 2 gender breakdown}
\small
      \begin{tabular}{p{0.35\columnwidth}p{0.40\columnwidth}p{0.06\columnwidth}}
      \midrule
         &  Women including trans women& 31\\
         &  Men including trans men & 30 \\
         & Non-binary & 0 \\
        \multirow{-4}{0.35\columnwidth}{\GreenUnderpaid} & Rather not to say & 0 \\
        \midrule
         &  Women including trans women& 45\\
         &  Men including trans men & 17 \\
         & Non-binary & 0 \\
        \multirow{-4}{0.35\columnwidth}{\YellowUnderpaid} & Rather not to say & 0 \\
        \midrule
        &  Women including trans women& 33\\
         &  Men including trans men & 27 \\
         & Non-binary & 0 \\
        \multirow{-4}{0.35\columnwidth}{\BlueUnderpaid} & Rather not to say & 2 \\
        \midrule
          &  Women including trans women& 39\\
         &  Men including trans men & 23 \\
         & Non-binary & 1 \\
        \multirow{-4}{0.35\columnwidth}{\PinkUnderpaid} & Rather not to say & 0 \\
         \hline
    \end{tabular}
    \label{tab:demo_study2}
\end{table}

\subsection{Participants}
Out of the target 280 participants, 22 experienced data logging issues, and 10 failed to be recruited due to platform monetary problems, resulting in 248 participants, with the following distribution across the four conditions: 63 for \PinkUnderpaid, 62 for \BlueUnderpaid, 62 for \YellowUnderpaid and 61 for \GreenUnderpaid. The distribution of participants' gender within these conditions is presented in \autoref{tab:demo_study2}. For each participant, we logged each completed step of the study, enabling us to perform an attrition analysis. 

In addition to these 249 participants, forty-eight additional participants dropped out of the study after consenting to participate; all did so during the explanation phase. Importantly, no indications of the conditions associated with the study were presented during this phase, suggesting that the dropout was not influenced by the perceived complexity of any specific condition.

\subsection{Analyses and results}

Our analysis is based on the data collected as reported in \cref{sec:expedesign}. 
In contrast to the findings from Study~1, a greater number of participants  (see \cref{fig:countgeneral}) allocated more money to rewarding performance (114,\minipie{46}{ColorPerfo}{\small 46\%}) than reducing the gender pay gap (97,\minipie{39.1}{ColorReduce}{\small 39.1\%}) or having no preference (37,\minipie{14.9}{ColorNoPref} {\small 14.9\%}). However, there was no difference in the amount of StudyCoin allocated for their preferred objective (39.62 CI[34.15; 45.10] for those who primarily reduced the gender pay gap and 37.91 CI[34.15; 42.47] 
for those who, in absolute value, put more money into rewarding performance; see \cref{sec:supplemental_materials} Supplemental Material).

An analysis of allocations, broken down by conditions, reveals that participants in the \GreenUnderpaid condition (see \cref{fig:count_conditions}) put more money on average towards reducing gender inequity (average of 12.6, CI[1.0; 23.9]) compared to the other three conditions in which participants slightly favored allocations rewarding performance. This drives the observed but small differences between the \nonstereotyped condition (\ie where participants could not infer the underpaid group correctly) and the \stereotyped condition. 


 

\subsection{Analysis by gender}\label{subsec:study2:ana-gender}

Analyzing the participant's allocation decisions broken down by condition and by gender revealed that participants in the \GreenUnderpaid condition were more likely to prefer reducing the pay gap regardless of their own gender, in line with our previous analysis. In contrast, in the remaining conditions, regardless of gender, participants were more likely to allocate more towards rewarding performance. In contrast to Study 1, women used the ``no preference'' strategy more than men (see \autoref{fig:count_gender_conditions}). More interestingly, compared to Study 1, in which participants, women in particular, used in-group strategies to allocate money, we did not find such an effect, contradicting our \textbf{HP3} (see \autoref{fig:count_gender_conditions} bottom). Indeed, women preferred to reward performance for all conditions, including the \PinkUnderpaid one in which employees' gender could have been inferred with color alone, and therefore we have no evidence to support \textbf{HP4}. The only trace of such an in-group effect can be found for men in both the \YellowUnderpaid and \PinkUnderpaid conditions (see \autoref{fig:count_gender_conditions} top), where men preferred rewarding performance and consequently were less like to reduce gender inequity. 

The analysis of the interaction between participants' gender and the two factors separately, \PGfactor and \Colorfactor, showed a very small but inconclusive difference for the \nonstereotyped factor (see \autoref{fig:count_gender_factors}). Finally, we note that the single non-binary participant in Study 2 allocated all funds towards reducing gender inequity.

\vspace{-0.5em}
\subsection{Qualitative analysis of open feedback}\label{subsec:study2:opentext}


In this study, we identified almost the same themes as in the first study.

\motherbox{
\AFbox{-2.8mm}{Two-step strategies}{ ``I first focused on reducing the gender pay gap to a reasonable change then allocated the rest to performance.'' (P65)}
\AFbox{-2.6mm}{Satisfaction}{``[I] am very satisfied with my allocation'' (P84).``I am extremely satisfied with the allocation.'' (P112)}
 \AFbox{-2.5mm}{Fairness}{
 ``Due to the current gender pay gap in order to ensure fairness I opted to give 40\% in reducing gender pay inequity'' (P91)}
 \AFbox{-2.8mm}{Need for more budget}{``I would've liked a higher budget because most of the budget had to go to adjusting gender inequality [...]'' (P105)}
}
As the explicit mention of gender (\eg the legend) was removed, the participants did not explicitly mention that the direction of the gender pay gap had influenced their decision.

Beyond the text analytics applied to Study 1, we additionally coded the responses focused on statements indicating participants' assumptions about gender-color associations. 
We observed that many participants attempted to attribute a specific gender category to a color. In other words, the non-presence of a legend added an ambiguity that participants addressed. This assignment was more expressed in the \stereotyped condition where participants associated pink with women and blue with men. 

\motherbox{\AFbox{-2.4mm}{\PinkUnderpaid}{ ``most of the higher perfomance is allocated to blue dot(men) and pink dots low perfomance therefore low earnings.'' (P91)}
 \AFbox{-2.4mm}{\BlueUnderpaid}{``Men seems to be higher performers but are being paid less than women.'' (P22)}
}

In the \nonstereotyped condition, fewer people mentioned an inferred attribution of gender with color. When they did, many participants attributed women to the underpaid group regardless of whether the underpaid group was in green or yellow, reflecting the societal norm of women being underpaid. 
\motherbox{
\AFbox{-2.4mm}{\GreenUnderpaid}{``There are fewer women in the grade C high performing group.'' (P106)
}
}

Interestingly, few people (and more in the \nonstereotyped condition) interpreted the gender categories by using color names instead of attempting to assign a color to a specific gender category. For those who did not try to assign a color to a specific gender, the use of \nonstereotyped colors can be useful to avoid in-group favoritism.

\motherbox{
\AFbox{-2.4mm}{\nonstereotyped}{``[...] Group A though in Group B C the \textcolor{W-NonStereo}{Olive} coloured gender appears to be slightly higher than the \textcolor{M-NonStereo}{Green}.'' (P190). }
}
\vspace{-1em}



\subsection{Discussion and summary}

With Study 2, we aimed to detect the effect of stereotyped color in the absence of other gender associations. 
Interestingly, our study did not detect clear differences between \stereotyped and \nonstereotyped colors---however, the allocation when green was underpaid was different from the case when blue was underpaid (and dissimilar from the cases when pink and yellow were underpaid, but the differences are were not large). Notably, removing the legend canceled the in-group favoritism effect found in Study 1. When we compared the allocations made by female participants in \PinkUnderpaid with \WLstereotyped in Study 1, we saw a large difference (see \autoref{fig:count_gender_conditions}, bottom).
The same observation holds for the \YellowUnderpaid condition and the \WLnonstereotyped in Study 1. When comparing the two studies regardless of participants' gender (see \autoref{fig:cis} and \autoref{fig:count_conditions}), the absence of the legend led to more decisions that rewarded performance over addressing inequity in all conditions except for \GreenUnderpaid. 

\section{General Discussion}

We evaluated the impact of \stereotyped colors on visualization-driven pay decisions. Through two crowdsourced user experiments, we examined the interactions between stereotyped gender data visualization and allocation decision-making. Our findings indicate that colors alone did not significantly influence pay decisions. However, our studies revealed other interesting insights. In particular, Study 1 showed that when gender was explicitly identified with a legend, there was a strong association between the participant's allocation decision, their own gender, and the gender of the underpaid group. In addition, many participants chose an allocation close to a fifty-fifty strategy, which they considered to be a ``fair'' allocation.

There are numerous factors that can affect a decision-making scenario such as ours, including fairness considerations, stereotype activation, in-group associations, and social norms. Below we discuss these factors and highlight avenues for future studies. We conclude this section with the potential limitations of a decision-making study in a crowdsourced environment and a reflection on our studies from the point of view of ethics in stereotype studies.

\subsection{Decision fairness, in-group dynamics and stereotypes}

Resource allocation represents a fundamental aspect of psychological inquiry, shaped by various principles that guide decision-making processes. Among these principles, fairness emerges as a tradeoff between different competing goals including two main strategies: equality (all individuals will obtain the same reward), and equity (individuals will be rewarded according to their contribution)
\cite{vanhootegemDifferentiatedDistributive2020,deutschEquityEquality1975,mannixEquityEquality1995}. Equality is often cited as the main principle used by decision-makers \cite{sampsonJusticeEquality1975,eekChoiceAllocation2009}, who then advocate for an equal distribution of shares among all recipients. In our studies, participants were not able to give each person exactly the same raise. Still, by allocating all the funds toward awarding performance, funds were attributed to everyone including the low performers, who would receive an allocation equal to 0.84\% of their pay. In contrast, an equitable allocation can be characterized as allocating all the funds to correcting the systematic underpayment of one group. Consequently, a fifty-fifty allocation strategy can be considered as an attempt to balance equality with equity. Such decisions may stem from a perceived sense of fairness where funds are spread across groups rather than targeted to maximize benefits for a sociodemographic or professional category (\ie gender, grade group, or performance level) \cite{eekChoiceAllocation2009,vandijk1995ASsymetricDilemmas}. However, importantly, any perception of fairness varies from one person to the next \cite{mannixEquityEquality1995}, and the association between the allocation decision and the perceived fairness of the allocation is a potential future research avenue within visualization-driven decision-making. 

Social Identify Theory states that people naturally categorize themselves and others into groups (in-groups = ``us'', out-groups = ``them'') and that they tend to favor their own group. \cite{Tajfel1986SocialIdentity}. In Study 1, women, on average, allocated more towards reducing the gender pay inequity when the gap negatively affected their in-group (\ie women) and on average allocated more towards rewarding performance when the inequity reduction favored their out-group (\ie men). This finding aligns with the literature \cite{eekChoiceAllocation2009,sanchez-mazasWhenOutgroup1994,cadsbyIngroupFavoritism2016}.
In Study 2, the absence of a legend impeded participants' ability to accurately identify their in-group, especially in the \nonstereotyped conditions. While some participants reported that they associated  
colors with gender categories, the lack of an explicit legend decreased the in-group effect but increased the ambiguity of the decision. In decision-making under uncertainty, decision-makers tend to be ambiguity averse (even choosing possibly worse but certain outcomes over ambiguous but possibly better outcomes) \cite{ camerer1992recent , ellsberg1961risk}. This ambiguity in turn may have led participants to prefer the certainty of the performance allocation over an ambiguous pay equity allocation. 

Social categorization of in-groups and out-groups is tied to stereotyping \cite{vanknippenbergSocialCategorization2000}. Intrinsically, data visualizations use pieces of information about phenomena or observations and group them into categories. Therefore, we argue that sociodemographic data visualization may lead to social categorization, but further studies are required to understand the potential detriments of such categorization in data visualization. We reason that in our studies social categorization has indeed occurred, especially in Study 1, and potentially and to a lesser extent in Study 2. A trigger must activate the categorization of someone or a group into a social category and, therefore, into an in-group or an out-group, and literature suggested that color may be such a trigger.

Applying stereotyped behaviors necessitates two distinct processes: stereotype activation and stereotype application \cite{kunda2003stereoactivation,Krieglmeyer2012ModelStereo}. Our experiments showed no definitive evidence that \emph{color alone} in data visualization was a cue strong enough to cause stereotype activation and its application. While existing literature in social cognition has established that colors can indeed trigger stereotyped behavior \cite{cunningham2011coloractivate}, it is crucial to note that the activation of stereotypes is not an automatic process. In addition, our studies' introduction and training phases were lengthy, around 15 minutes, compared to the decision task (approximately one minute). Much of study participant's time was dedicated to reading detailed explanations regarding the study's various features. We designed these explanation and training phases to ensure that participants could perform the task, accounting for varying data visualization literacy levels and non-native English speakers. This may have introduced a high cognitive load for those specific participants. As demonstrated by the literature \cite{Bodenhausen1998StereotypeInhibition,spencerAutomaticActivation1998,gilbertTroubleThinking1991}, a high cognitive load \emph{before} an exposition of the stereotype, in our case, the data visualization used to perform the allocation, can explain the non-activation of the associated stereotypes. Our study results might be explained by this literature, highlighting the non-automatic nature of stereotype application even in the presence of a stereotyped trigger under a high cognitive load. Interestingly, in Study 2, some participants assigned a color to a specific gender group (as discussed in \cref{subsec:study2:opentext}) even without a legend and even when \nonstereotyped colors were used.  Therefore, the potential for the activation and application of stereotypes in visualizations warrants further discussion and study.

Our studies highlight the complexity of studying stereotyping for visualization-supported decision-making. Studying stereotyping is a meticulous task in which many users' characteristics need to be understood and controlled for. And even though data visualization has components that can trigger or induce stereotypes, such as colors in our study, the evidence that data visualization can activate a stereotype and impact decision-making is still inconclusive. Indeed, the abstract nature of data visualization, coupled with the fundamental exploratory and analytical objectives of visualizations in a decision-making context, may protect users from such stereotype-induced biases. In addition, in our study, the stereotypes interacted with the social norm of women’s underpayment, which for some may have served as a decision justification. Finally, the objective of avoiding stereotypes by making careful decisions, which has been identified as a key objective in HR decision-making \cite{Koch2015MetaGender}, may also have played a role. Thus, further investigation is needed to disentangle the relationship between stereotypes, data visualization, social norms, objectivity training, and decision-making.

\subsection{Crowdsourced evaluations of decision-making}

Various crowdsourcing platforms have emerged over the past two decades, allowing researchers to recruit a large sample of participants that share some characteristics. Data visualization researchers have increasingly utilized these crowdsourced platforms (as documented in the following review \cite{borgoInformationVisualization2018}). However, it is still unclear what types of tasks lead to similar results in a lab environment. Notably, Kim and colleagues \cite{kimInvestigatingEfficacy2011} could not replicate a decision-making study made in a laboratory in a crowdsourced environment, identifying participants' reliability as a significant challenge, particularly with respect to mitigating the impact of ``poor'' participants. The introduction of attention checks, comprehension tests, and/or time-restricted tasks has somewhat alleviated these issues. 

Yet, in our two studies, we observed high levels of engagement by participants. Three of the four open-ended questions were optional (although nothing was present on the screen to indicate they were optional). Most participants decided to spend time answering them. In addition, many of their answers aligned with the questions asked, indicating that participants were engaged with the open-text questions. However, we also identified that some ASI scale answers showed little motivation at the end of the study, possibly reflected in the insignificant associations between the ASI scale scores and the allocation decisions. 

Despite the advantages presented by crowdsourcing platforms, research into the phenomena of stereotyping and decision-making should also be compared to studies with more interactive engagement between participants and the examiner. In our studies, while a significant number of participants contributed more than the requisite 100 characters for the single mandatory questions, the impossibility of looking deeper into their thoughts during a formal interview phase presents a notable limitation. This absence of qualitative interaction restricts a deeper understanding of their thought patterns and decision-making rationales. Future research on how to evaluate visualization-driven decision-making in crowdsourced experiments is crucial for researchers and practitioners to know what works and what doesn't.


\subsection{A short ethical reflection}
{\setlength{\spaceskip}{0.22em plus 0.05em minus 0.02em }
Studying stereotypes requires the presentation of a stimulus to activate the stereotypes and, in the end, asking participants to perform a task that may be altered by the application of these stereotypes. While such methodologies are essential to studying the stereotypes, they inherently carry the risk of perpetuating and/or reinforcing the stereotypes  \cite{cecchiniReinforcingReproducing2019}. In our context, this reinforcement could take the form of reinforcing misconceptions surrounding the gender pay gap. In our task, the gender pay gap starts at 5\% and cannot be reduced below 1\%; some participants may therefore think we consider 1\% as acceptable---which we do not. 

As Claartje Vinkenburg \cite{Vinkenburg2017InterventionMitigate} has described, it is possible for numerous interventions in institutions initially aiming to reduce inequity to ultimately turn into strong resistance and, for some, a backfire effect. In a plea for better inclusion of ethical concerns when studying stereotypes \cite{cecchiniReinforcingReproducing2019}, Mathilde Cecchini proposed three guidelines to reduce the potential reinforcement of a stereotype: cultivating an ethical sensibility, constructing and posing adequate questions and facilitating reflections. Unfortunately, these guidelines are primarily tailored to in-person studies, where direct interaction with participants is more feasible.

Therefore, in our studies, we limited the potential reinforcement of stereotypes by presenting gender as a variable like any other and asking participants to tell us, in the post-task questions, what they thought about the decision they just made.

\section{Conclusion}

In this paper, we present an investigation of the use of \stereotyped colors for pay decisions.
In a series of two crowdsourcing experiments (N= 276, N = 248) with domain experts, we evaluated how the use of \textcolor{W-Stereo}{pink} and \textcolor{M-Stereo}{blue} compared with \nonstereotyped colors, affected how participants allocated limited funds between reducing a gender pay gap in their company and rewarding employees for their performance. These studies show the entanglements of stereotyping, decision-making, and individual characteristics, such as the decision-maker's gender and prior knowledge. We see this work as a first brick in the foundational understanding of the effect of social categorization such as stereotyping in visualization-supported decision-making.

\section*{Supplemental Materials}
\label{sec:supplemental_materials}
All supplemental materials are available on OSF at \url{https://osf.io/d4q3v/?view_only=22b636d6f7bb4a7991d9576933b3aaad}, which include (1) the data from the two studies, (2) a demo video of the two studies, (3) the source code for the website hosting the two studies, (4) the scripts to analyze the data and generate the figures, (5) all figures from this paper and (6) additional figures.

\bibliographystyle{abbrv-doi-hyperref}

\bibliography{mybib}




\end{document}